# An Effective Reduction of Critical Current for Current-Induced Magnetization Switching by a Ru Layer Insertion in an Exchange-Biased Spin-Valve

Y. Jiang, S.Abe, T.Ochiai, T.Nozaki, A.Hirohata, N.Tezuka and K.Inomata

Department of Materials Science, Graduate School of Engineering, Tohoku University, and CREST, Japan Science and Technology Agency, Sendai 980-8579, Japan

Recently it has been predicted that a spin-polarized electrical current perpendicular-to-plane (CPP) directly flowing through a magnetic element can induce magnetization switching through spin-momentum transfer. In this letter, the first observation of current-induced magnetization switching (CIMS) in exchange-biased spin-valves (ESPVs) at room temperature is reported. The ESPVs show the CIMS behavior under a sweeping dc current with a very high critical current density. It is demonstrated that a thin Ruthenium (Ru) layer inserted between a free layer and a top electrode effectively reduces the critical current densities for the CIMS. An "inverse" CIMS behavior is also observed when the thickness of the free layer increases.

Giant magnetoresistance (GMR) was first found in Fe/Cr multilayers and later observed in a large number of transition metal based multilayers due to spin-dependent scattering at the nonmagnetic/magnetic interfaces [1]. In 1996, Slonczewski and Berger predicted magnetization reversal induced by spin-polarized current injection in current-perpendicular-to-plane (CPP) magnetic multilayers through spin-transfer effect [2]. It is very interesting because current-induced magnetization switching (CIMS) instead of magnetic field switching can offer a completely new class of "current-controlled" memory devices [3]. For example, it can greatly simplify the design of high-density nonvolatile memory. Theoretical [4-6] and experimental [7-12] understandings about the CIMS effect have been primarily established. But the basic physical processes involved in the switching are not yet fully understood. Till now almost all of the reported experimental works have been concentrated on the CIMS observed in nanometer-sized Co/Cu/Co or other pseudo spin-valves (SPVs) pillars. It would therefore be a breakthrough if the CIMS phenomenon is observed in exchange-biased SPVs (ESPVs) because the ESPVs are always used in the magnetic storage devices instead of simple ferromagnet (FM)/Cu/FM trilayers in order to precisely control their domain structures and to decrease their coercivities. In this letter, we report the first observation of the CIMS in ESPV nanofabricated pillars, consisting of an antiferromagnet (AF)/pinned FM layer/Cu/free FM layer/(Ru cap) structure, which can stand the injection of high dc current density (more than $10^9$ A/cm$^2$). The average critical current density for the CIMS in the ESPVs is around $1.6\times10^8$ A/cm$^2$, much higher than the reported values in Co/Cu/Co nanopillars [9,11,12]. A thin Ru cap layer (0.45 nm) dramatically decreases the average critical current density down to $2.2\times10^7$ A/cm$^2$ when the free layer is kept as 2.5 nm. For those ESPVs with a thick free layer (5 nm), we observe an "inverse" CIMS behavior — the injected current flowing from the free to the pinned FM layers favors an antiparallel (AP) magnetic configuration, while the current flowing from the pinned to the free layers favors a parallel (P) configuration, which is the opposite of the reported normal CIMS behaviors.

The structure of our samples was Cu (20 nm)/IrMn (10 nm)/$Co_{90}Fe_{10}$ (5 nm)/Cu (6 nm)/ $Co_{90}Fe_{10}$ ($d_1$ nm)/Ru ($d_2$ nm)/Cu (5 nm)/Ta (2 nm). The $Co_{90}Fe_{10}$ free layer thickness $d_1$ was varied between 2.5 and 5 nm. The thickness of the Ru layer was changed between 0 and 0.45 nm. The fabrication process was "subtractive" and was optimized carefully. The multilayer was first deposited on a Si/$SiO_2$ substrate in an ultrahigh-vacuum sputtering system with a base pressure of below ~$5\times10^{-9}$ Torr. A 200 Oe magnetic field was applied in order to induce an easy axis during the sputtering. Then Cu bottom electrodes and Cu/Ta top electrodes were patterned using electron-beam lithography and subsequent ion milling etching. A GMR ESPV element was etched out followed by $SiO_2$ sputtering. A thick Cu top layer was then deposited using a lift-off process. Figure 1(a) shows a scanning electron microscopy (SEM) image of the top view of an ESPV pillar. The picture is taken before sputtering $SiO_2$. For the pillar in Fig. 1 (a), the area $A$ of the nanopillar is measured to be $280\times90$ nm$^2$. It should be noted that all the sizes of the samples mentioned in this paper are the measured values. A cross-sectional schematic diagram of the final patterned device is shown in Fig. 1(b). As shown in Fig. 1(b), the conventional four probes measurement of transport properties is carried out in CPP geometry at room temperature with a magnetic field applied along the easy axis. In our system, positive current flows from the free to the pinned FM layer.

When the free layer thickness $d_1$ is 2.5 nm and the Ru cap layer thickness $d_2$ is 0 nm (i.e. no Ru cap layer), the measured CPP GMR loop of one ESPV pillar ("ESPV 1") with a size $A$ of 280×90 $nm^2$ is shown in the upper left inset of Fig. 2. The CPP GMR of ESPV 1 is around 0.3%. We also measure the resistance switching while sweeping a dc current. We start the measurement while keeping the magnetic field as 90 Oe that makes the AP state stable in the sample. The sweeping dc current does cause the resistance switching in ESPV 1, shown as the closed circles with lines in Fig. 2. The hysteretic CIMS loop is very similar to the previously reported data on the Co/Cu/Co nanopillars [9,11,12]. At the beginning, the resistance keeps nearly constant with increasing positive current. When the positive current reaches a critical value $I_P$~27 mA (i.e. the critical current density $J^{AP\rightarrow P} = 1.1\times10^8$ A/cm$^2$), the resistance jumps to a lower value, which corresponds to the resistance in the P configuration. When the current decreases to a critical value $I_{AP}$~-54 mA (i.e. the critical current density $J^{P\rightarrow AP} = -2.1\times10^8$ A/cm$^2$), the resistance jumps back to the higher value, the AP resistance. We define the average critical current density as $J_C=1/2(|J^{AP\rightarrow P}| + |J^{P\rightarrow AP}|)$. The calculated $J_C$ for ESPV 1 is $1.6\times10^8$ A/cm$^2$. The abrupt change in resistance by the injected current is the same as the observed CPP GMR in an external magnetic field. This means that the resistance switching by the injected dc current in ESPV 1 is due to the full reversal of magnetization of the sample through the spin-transfer effect. For ESPV 1, the critical current $|I_P|$ is lower than $|I_{AP}|$, which is in good agreement with the spin-transfer model [2]. We note that the critical current densities here are much higher than the reported ones in the Co/Cu/Co nanopillars [9], typically in the order of ~$10^7$ A/cm$^2$. The thick AF (~10 nm) in ESPV1 may cause a strong depolarization of the spin current. The reduced spin-polarization leads to even smaller spin-torque per unit current applied to the FM free layer and therefore a large current is required to drive the CIMS.

For the sample "ESPV 2", we keep the free layer thickness as $d_1$=2.5 nm and insert a thin Ru layer ($d_2$=0.45 nm) between the free layer and the top Cu layer. The junction size $A$ of ESPV 2 is also 280×90 nm$^2$. The I-R curve measured under a 90 Oe magnetic field together with its GMR loop are shown as the open circles with line in Fig. 2 and the inset. A clear resistance switching by an injected dc current can be observed. The CIMS behavior is similar to that of ESPV 1 except that the average critical current density is much lower, 2.2×10$^7$ A/cm$^2$. Therefore by inserting a thin Ru cap layer between the free layer and the top electrode, the average critical current density that makes the full reversal of magnetization is effectively decreased.

In the report by Urazhdin *et al.* [12], a strong spin-scatterer $Fe_{50}Mn_{50}$ layer being inserted between a nanopillar and a top electrode can also decrease the critical current density. They argue that the insertion of $Fe_{50}Mn_{50}$ layer increases the spin-polarization inside the nanopillar and therefore enhances the spin-transfer effect. We believe that there exists a similar story in our system for the ESPVs with the thin free layers. It has been theoretically predicted that Ru impurity in Co scatters majority spins more strongly and leads to large spin imbalance and spin scattering at a Co/Ru interface [13]. In our system, the thin Ru cap serves as a strong majority spin-scatterer. In fact, we have observed CPP GMR enhancement in the ESPVs with a thin Ru cap structure [14] that may due to the large Co/Ru interface scattering and increased spin-polarization inside the ESPVs by the Ru cap layer. It is noteworthy that the ratio $|J^{P\to AP}|/|J^{AP\to P}|$ deceases from ~1.9 for ESPV 1 to ~1.5 for ESPV 2. Therefore the Ru cap in the ESPVs has similar effect as the thickness reduction of the Cu space layer in Co/Cu/Co sandwich structure, i.e. enhancement of the spin-polarization [15]. According to Urazhdin *et al.*, in the spin-transfer process, the spins opposite to the free layer moment $M$ generate magnetic excitations when they flip their spins, while those spins along $M$ absorb the excitations when they spin-flip, therefore the CIMS is expected to be determined by the spin-polarization. A large spin-polarization leads to large spin-torque per unit current and therefore reduces critical current. Fig. 3 gives the resistance change $\Delta R$ dependence of $1/J^{P\rightarrow AP}$ and $1/J^{AP\rightarrow P}$ for the ESPVs with different thickness of the Ru cap. For comparison, the results by Urazhdin *et al.* [12] have also been shown in the figure. Similar to the normal SPVs, the ESPVs also show linear dependence between the inverse of the critical current density and the $\Delta R$, except that the gradient of the linear line for ESPVs is lower than that for normal SPVs.

From the point of view of spin-accumulation, the insertion of a thin Ru scatterer between nanopillar and top electrode serves as a "wall" of majority spins and reduces the spin accumulation outside a nanopillar. Therefore the structure permits higher spin accumulation inside the nanopollar. Spin-torque is proportional to the transverse spin accumulation (in the plane transverse to the magnetization of magnetic layer) [16, 17]. The enhancement of the transverse spin accumulation increases the spin-torque and therefore effectively reduces the critical currents. The enhanced spin-torque can easily be understood by the calculation of spin-transfer efficiencies for both ESPV 1 and ESPV 2. In a multilayer, the resistance change at the critical current $I_C$ is because of the onset of coherent magnetization precession and spin wave generation in the magnetic free layer.

Based on the Landau-Lifshitz-Gilbert (LLG) equation, the critical current has an expression [18]:

$$I_C = \frac{et_{free}}{\hbar\varepsilon}[\frac{23.4M_S D}{2\hbar\gamma} + 6.3r^2\alpha_{LLG}M_S(H_{app} + H_{ex} - M_{eff})]. \quad (1)$$

Here $t_{free}$ is the thickness of the free layer, $M_S$ the saturation magnetization, $r$ the contact radius, $\gamma$ the gyromagnetic ratio, $D$ the spin wave exchange stiffness, $\alpha_{LLG}$ the LLG damping parameter, and $\varepsilon$ the spin-transfer efficiency. $H_{app}$ is the applied magnetic field, $H_{ex}$ the effective interlayer exchange field, and $M_{eff}$ the effective saturation magnetization including surface anisotropy. The CIMS measurements shown in Fig. 2 are always carried out under the application of an external magnetic field. Thus here we simply assume $H_{app} = M_{eff} - H_{ex}$. From Eq. (1) we obtain $I_C \sim \frac{1}{\varepsilon}$, i.e. the critical current is inversely proportional to the spin-transfer efficiency $\varepsilon$. We use $J_{C1}$ and $J_{C2}$ to represent the average critical current densities for ESPV 1 and ESPV 2, respectively. $\varepsilon_1$ and $\varepsilon_2$ are the spin-transfer efficiencies in ESPV 1 and ESPV 2, respectively. Then we obtain

$$\frac{\varepsilon_2}{\varepsilon_1} = \frac{J_{C1}}{J_{C2}} = \frac{1.8\times10^8 A/cm^2}{2.2\times10^7 A/cm^2} \approx 8.2\,. \quad (2)$$

Equation (2) means that the Ru cap layer greatly increases the spin-transfer efficiency (about 8.2 times) compared with the ESPVs without the Ru cap. If $\varepsilon_2$ has an ideal value of 0.5, $\varepsilon$ is estimated to be only 0.06. The low spin-transfer efficiency can explain why it is so difficult to achieve the CIMS in an ESPV.

For further study of the CIMS properties of ESPV, we double the free layer thickness $d_1$ into 5 nm. The thickness $d_2$ of the Ru layer is 0.45 nm. The pillar area $A$ is maintained as 280×90 nm$^2$. We surprisingly obtain an “inverse” CIMS behavior in all the five samples studied. The results of one sample "ESPV 3" is shown in Fig. 4. After increasing and keeping the magnetic field as 250 Oe, i.e. starting from point "A", the injected dc current causes a resistance change that is corresponding to the magnetoresistance. However, in this sample, i.e. the ESPV with a thicker free layer, a certain positive current drives P to AP, while a negative one drives AP to P, which is the opposite of the conventional spin-transfer effect as shown in Fig. 2. In our system, the positive current means that the electrons flow from the pinned to the free FM layer, so the "inverse" CIMS in Fig. 4 is probably because of the magnetization switching of the bottom FM layer. Usually it is difficult for the bottom FM layer to be switched because it is "pinned" by the AF layer. But when both the pinned and free FM layers possess the same thickness and the external magnetic field is as high as 250 Oe, which is close to the reversal field of the pinned layer, it should be reasonable to assume the switching of the pinned FM layer.

In summary, the effects of the thin Ru cap layer insertion on the CIMS behavior of the ESPVs are studied at room temperature. For the ESPVs with the thin free layers of 2.5 nm, the Ru cap layers effectively decrease the critical current density. For the ESPVs with the thick free layers of 5 nm, an "inverse" CIMS behavior has been observed. Combining the spin-transfer and the reflection of majority spins by the Ru layer, the observed effects of the Ru layer on the CIMS can be understood. We believe that our first observation of the CIMS in the ESPVs and the efficient decrease of the critical current density by inserting the thin Ru cap layers will excite more studies on the related structures and expedite the steps to future "current-controlled" magnetic storage and sensing devices.

We would like to thank Prof. Jack Bass and Dr. Shufeng Zhang for the helpful discussions. This work was supported by IT-program of Research Revolution 2002 (RR 2002), the Priority Area, 1407602 from MEXT, and NEDO grant for NAME and SCAT.

**Figure captions:**

Fig. 1. (a) SEM image of the top view of an ESPV pillar. (b) Schematic cross-sectional diagram of a patterned CPP ESPV pillar with top and bottom electrodes.

Fig. 2. Resistance (R) vs sweeping dc current (I) for ESPV 1 (closed circles with lines) and ESPV 2 (open circles with lines). Both curves have been measured under a 90 Oe magnetic field. The insets show the GMR loops of ESPV 1 (closed circles with lines) and ESPV 2 (open circles with lines).

Fig. 3. The resistance change ΔR dependence of $1/J^{P\rightarrow AP}$ (closed upward triangles) and $1/J^{AP\rightarrow P}$ (closed downward triangles). For comparison, the results by Urazhdin *et al* [12] have also been shown (open triangles) in the figure. The linear lines (both solid and dash) are guide to eyes.

Fig. 4. I-R and GMR curves for ESPV 3. The I-R curve has been measured under a 250 Oe magnetic field and the dc current has been swept along the direction from 1 to 6.

Fig 1 (Y.Jiang et al.):

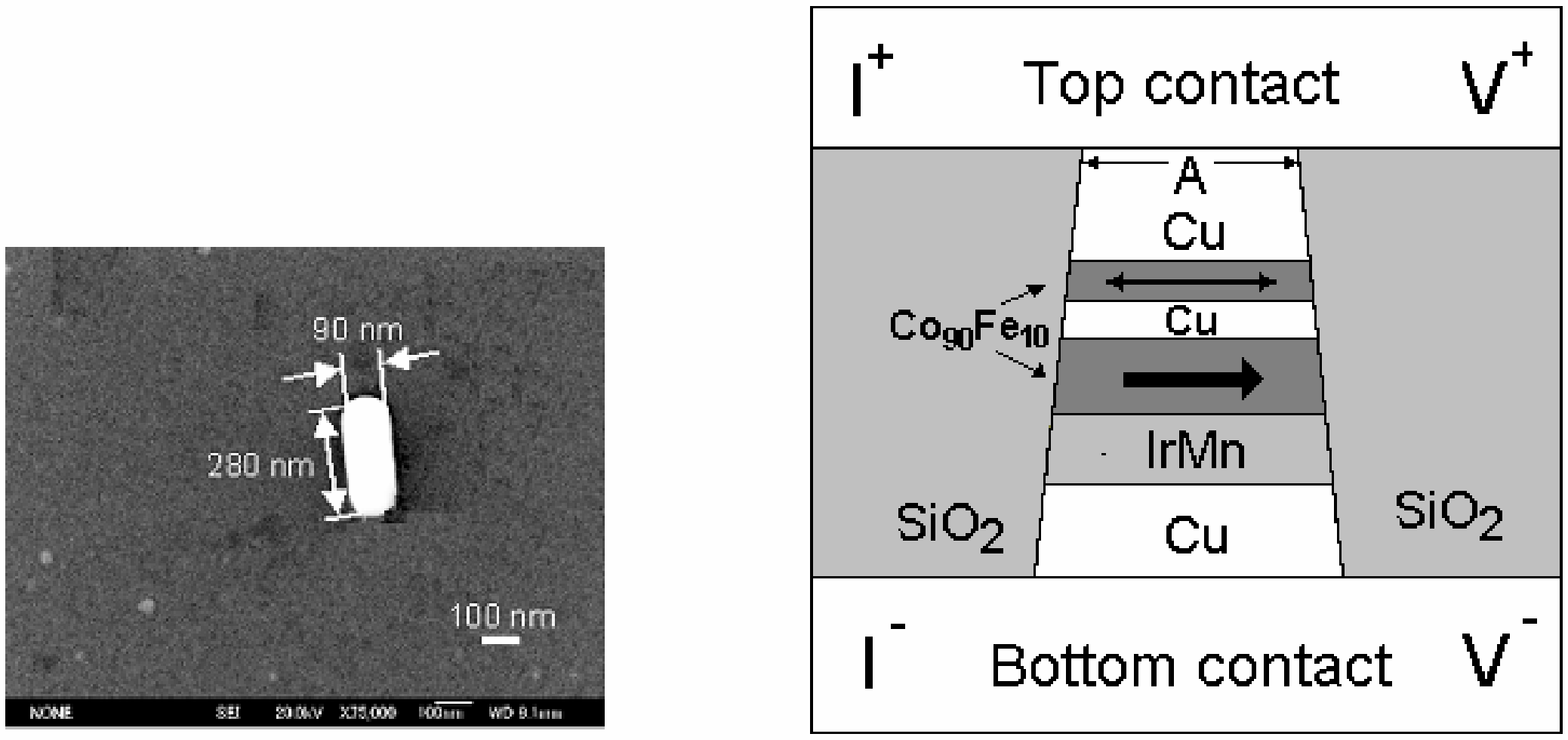

(a) (b)

Fig 2 (Y.Jiang et al.):

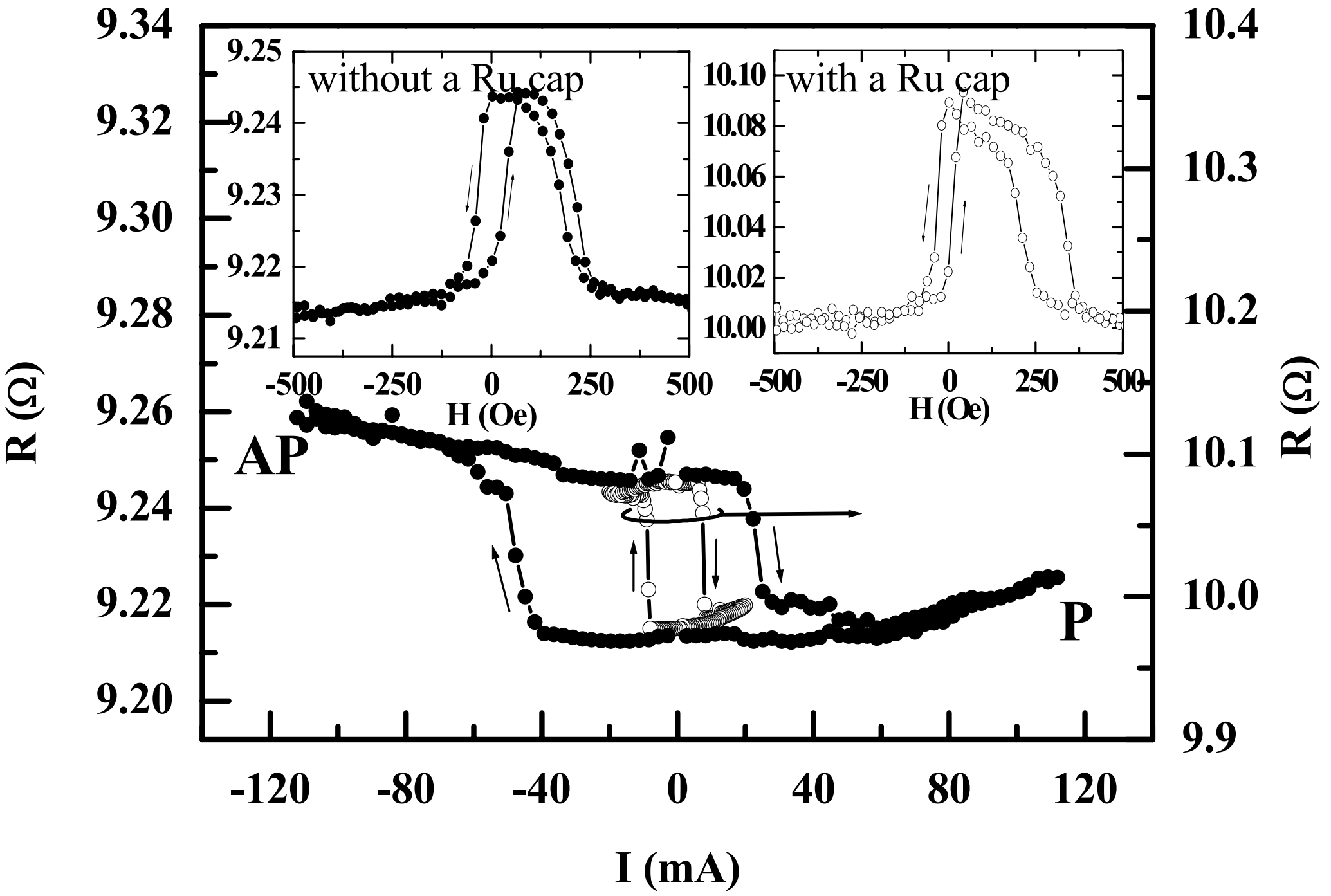

Fig 3 (Y.Jiang et al.):

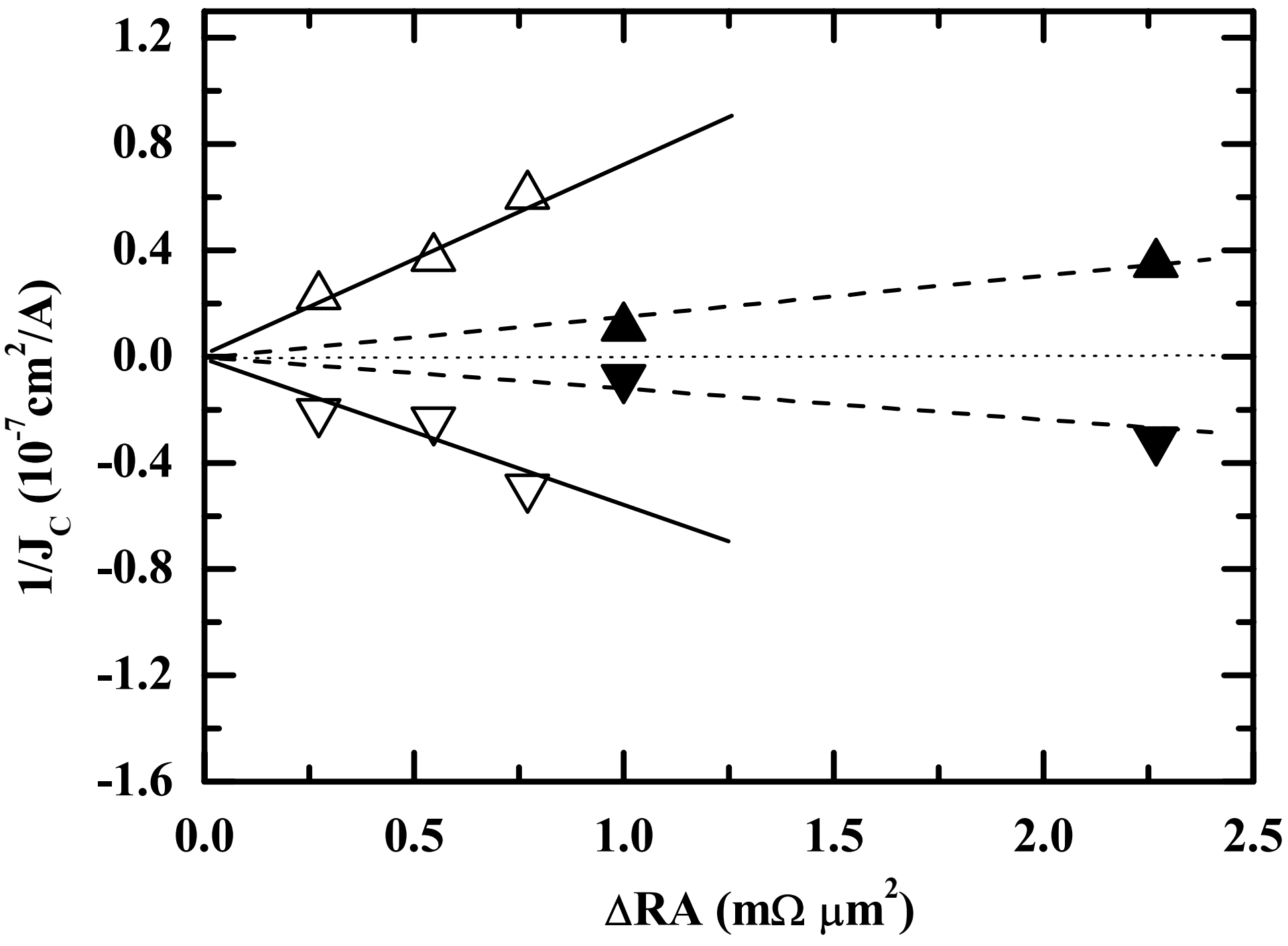

Fig 4 (Y.Jiang et al.):

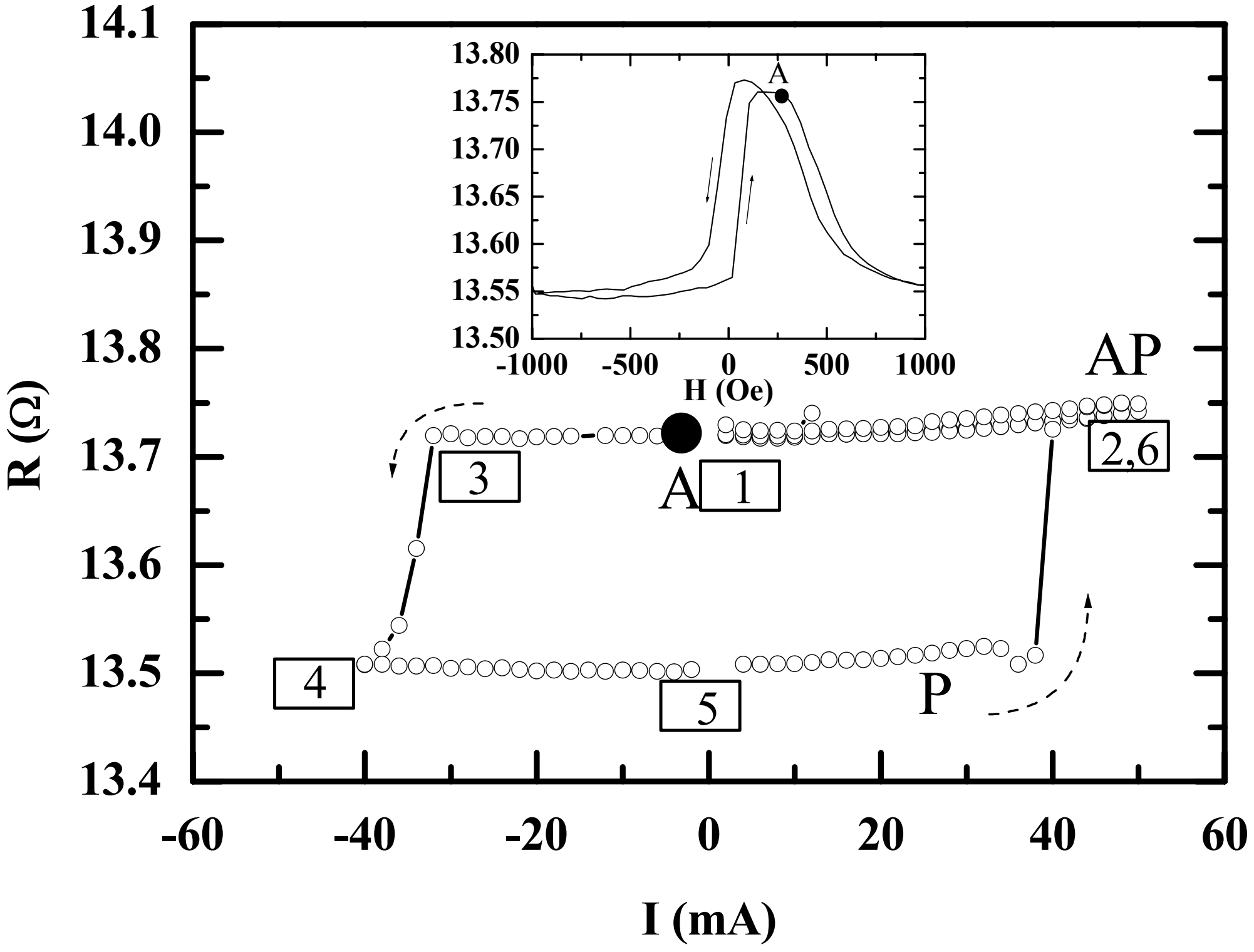